\documentclass[prd,twocolumn,superscriptaddress,altaffilletter,amssymb,showpacs,nofootinbib]{revtex4}

\usepackage[dvips]{graphicx}
\usepackage{amsmath}
\newcommand{\be}{\begin{equation}}
\newcommand{\ee}{\end{equation}}
\newcommand{\bea}{\begin{eqnarray}}
\newcommand{\eea}{\end{eqnarray}}
\newcommand{\vphi}{\varphi}

\begin{document}

\title{Study Of Tachyon Dynamics For Broad Classes of Potentials}

\author{Israel Quiros}\thanks{iquiros@fisica.ugto.mx}\address{Divisi\'on de Ciencias e Ingenier\'ia de la Universidad de Guanajuato, A.P. 150, 37150, Le\'on, Guanajuato, M\'exico.}

\author{Tame Gonzalez}\thanks{tame@uclv.edu.cu}\address{Departamento de F\'isica, Universidad Central de Las Villas, 54830 Santa Clara, Cuba.}

\author{Dania Gonzalez}\thanks{dgm@uclv.edu.cu}\address{Departamento de Matem\'atica, Universidad Central de Las Villas, 54830 Santa Clara, Cuba.}

\author{Yunelsy Napoles}\thanks{yna@uclv.edu.cu}\address{Departamento de Matem\'atica, Universidad Central de Las Villas, 54830 Santa Clara, Cuba.}

\author{Ricardo Garc\'ia-Salcedo}\thanks{rigarcias@ipn.mx}\address{Centro de Investigacion en Ciencia Aplicada y Tecnologia Avanzada - Legaria del IPN, M\'exico D.F., M\'exico.}\author{Claudia Moreno}\thanks{claudia.moreno@cucei.udg.mx}\address{Departamento de F\'isica y Matem\'aticas, Centro Universitario de Ciencias Ex\'actas e Ingenier\'ias, Corregidora 500 S.R., Universidad de Guadalajara, 44420 Guadalajara, Jalisco, M\'exico.}

\date{\today}

\begin{abstract}

We investigate in detail the asymptotic properties of tachyon cosmology for a broad class of self-interaction potentials. The present approach relies in an appropriate re-definition of the tachyon field, which, in conjunction with a method formerly applied in the bibliography in a different context, allows to generalize the dynamical systems study of tachyon cosmology to a wider class of self-interaction potentials beyond the (inverse) square-law one. It is revealed that independent of the functional form of the potential, the matter-dominated solution and the ultra-relativistic (also matter-dominated) solution, are always associated with equilibrium points in the phase space of the tachyon models. The latter is always the past attractor, while the former is a saddle critical point. For inverse power-law potentials $V\propto\phi^{-2\lambda}$ the late-time attractor is always the de Sitter solution, while for sinh-like potentials $V\propto\sinh^{-\alpha}(\lambda\phi)$, depending on the region of parameter space, the late-time attractor can be either the inflationary tachyon-dominated solution or the matter-scaling (also inflationary) phase. In general, for most part of known quintessential potentials, the late-time dynamics will be associated either with de Sitter inflation, or with matter-scaling, or with scalar field-dominated solutions.

\end{abstract}


\maketitle

\section{Introduction}

Inflationary models of the universe have been studied from the string theory perspective because inflation provides an explanation for the homogeneity and isotropy of the early universe. Additionally, recent astrophysical observations indicate us that the universe is presently undergoing a phase of accelerated expansion that has been attributed to a peculiar kind of source of the Einstein's field equations acknowledged as ``dark energy'' \cite{acce-exp}.\footnote{For an extensive review see \cite{DDE,PCA}.} The crucial feature of the dark energy which ensures an accelerated expansion of the universe is that it breaks the strong energy condition. The tachyon field arising in the context of string theory \cite{7} provides one example of matter which does the job. This is one of the reasons why the tachyon has been intensively studied during the last few years in application to cosmology both as a source of early inflation and of late-time speed-up of the cosmic expansion \cite{tachy cosmo}-\cite{Copeland}. The proposals where the tachyon field acts as a source of dark energy are highly correlated with specific forms of the self-interaction tachyon's potential \cite{11,16,aguirre-lazkoz,Copeland}.

Type II string theories have two kind of D-branes: BPS and non-BPS D-branes. The former ones are stable solitons which break half of the spacetime supersymmetries, while the latter ones suffer from open string tachyonic mode whose mass causes the brane to be unstable in a flat background. Decay of unstable D-branes is an interesting process which has shed some new light in understanding properties of string theory in time-dependent backgrounds. An effective action of the Dirac-Born-Infeld type proposed in \cite{garousi} can capture many properties of these decaying processes. The system considered is a superstring theory compactified on a six dimensional compact manifold so that there are 3 large spatial directions. All the moduli are assumed to be frozen. In this theory an unstable D-brane extends along the three large spatial directions. This system is modeled by the following effective action \cite{tachy cosmo,garousi}: 

\bea &&S=\frac{1}{2}\int d^4x\sqrt{|g|}R+S_\vphi,\nonumber\\&&S_\vphi=-\int d^4x\sqrt{|g|} V(\vphi)\sqrt{1+(\partial\vphi)^2},\label{sen}\eea with $\vphi$- the tachyon scalar field, $V$-its self-interaction potential, and $(\partial\vphi)^2\equiv g^{\mu\nu}\partial_\mu\vphi\partial_\nu\vphi$. The tachyon potential $V(\vphi)$ has a maximum at $\vphi=0$ and a minimum at $\vphi=\vphi_0$ where it vanishes. Since the minimum of the potential describes a configuration where there are no D-branes \cite{sen1}, around this minimum there are no physical open string excitations. If one lets the tachyon roll beginning with any spatially homogeneous initial configuration, it evolves asymptotically towards its minimum instead of oscillating about the minimum \cite{roll tachy}. The total energy density is conserved during the evolution but the pressure evolves to zero \cite{FT of TM}. 

The effective action (\ref{sen}) has become popular in cosmological applications as the starting point to study the dynamics of tachyon fields. In the reference \cite{tachy cosmo}, for instance, the author investigates the cosmic dynamics of a universe described by a decaying (unstable) D-brane, where a specific form of the tachyon potential $\propto\cosh^{-1}$ has been considered. A inflationary solution was found, however, the inflation produced by the solution is not slow roll, so that this solution cannot be used for a realistic model of primordial inflation. Nonetheless such solutions could have played a role in the early history of the universe, e.g. during a preinflationary phase \cite{tachy cosmo}. As already mentioned at the beginning of this introductory section, other kinds of potentials can lead to other kinds of cosmic dynamics, so that the above effective theory of the tachyon field (\ref{sen}) has been used also to describe the present period of speed-up of the cosmic expansion (see \cite{11,16,aguirre-lazkoz,Copeland} for an incomplete list of references).

A dynamical systems study of Friedmann-Robertson-Walker (FRW) cosmology within phenomenological theories based on the effective tachyon action (\ref{sen}) can be found in Ref.\cite{Copeland}. However the authors of \cite{Copeland} were able to study self-interaction potentials of the power-law type only. For more general potentials the corresponding system of ordinary differential equations in the phase space is not a closed system of equations any more, and one has to rely on the notion of ``instantaneous critical points'' whose physical relevance is unclear.

Due to the importance of investigating the eventual impact of a primordial tachyon field of fundamental origin\footnote{In the present models the tachyon field is considered as an open string mode which causes the D-brane where he lives to be unstable.} on the cosmic dynamics (anticipated by Sen in \cite{tachy cosmo}), including possible signature left by these field in present-day cosmology \cite{11,16,aguirre-lazkoz,Copeland}, and due to the strong dependence of the tachyon field dynamics on the specific form of the self-interaction potential, in this paper we aim at studying the cosmological dynamics of the tachyon model given by the effective action (\ref{sen}), for a broad class of potentials. To this end we shall perform an appropriate transformation of the tachyon field, which, in combination with the application of an approach formerly used in a different context, for instance in Ref.\cite{chinos}, allows to study a vast variety of self-interaction potentials. The present approach also enables to correlate the results of dynamical systems studies of certain generalized Dirac-Born-Infeld (DBI) models \cite{speedlimit}-\cite{DBInflation} already existing in the bibliography, with apparently unrelated results of studies of tachyon field cosmology. We will be able, in particular, to identify the results of Ref.\cite{Copeland} within the context of tachyon cosmology, with those in \cite{copeland-shuntaru} for a DBI-field with exponential potential and brane tension, not reported previously.

The paper has been organized in the following manner. The mathematical aspects of the cosmological model we will be focusing on, as well as the field equations of the transformed theory, are given in section II. Section III is devoted to the study of the asymptotic properties of the latter transformed model for a generic types of self-interaction potentials, through the application of the dynamical systems tools. In section IV we discuss the relevant aspects of the dynamics of the transformed model whose dynamics has been discussed in the previous section, and of the corresponding untransformed tachyon model, so that the results of section III can be safely translated to the case of interest (tachyon cosmology). It will be demonstrated, also, that former studies within the phenomenological tachyon model given by the effective action (\ref{sen}), that were constrained to power-law potentials only \cite{Copeland}, can be generalized to exponential type of potentials. A detailed discussion of the main results of the paper is presented in section V, while the conclusions are given in section VI. For self-consistency and completeness, an appendix with the basic recipes of the application of the dynamical systems tools to cosmology has been added. In this paper we will be focusing in homogeneous Friedmann-Robertson-Walker cosmological space-times. We use natural units $c=8\pi G=1$.

\section{Basic Equations and Set-Up}

The phase-plane analysis of tachyon dynamics within FRW cosmology has been extensively studied in \cite{aguirre-lazkoz}. Most part of these studies have been performed by considering only the inverse square tachyon potential \cite{10,11,aguirre-lazkoz,Copeland,square}, as long as only the inverse square potential allows to obtain a closed autonomous dynamical system out of the evolution equations. For any other potentials the number of phase-space dimensions will be higher if the system of ordinary differential equations (ODE) is to remain a closed one as we will check in the next section. To go beyond the square-law tachyon potentials requires to rely on the notion of "instantaneous critical point" whose meaning from the point of view of the dynamical systems is unclear \cite{Copeland}. 

In this section we will propose a way out of this problem. To that purpose we shall transform the tachyon field $\vphi\rightarrow\phi$ as it follows:

\be \vphi\rightarrow\phi=\int d\vphi\sqrt{V(\vphi)}\;\Leftrightarrow\;\partial\vphi=\frac{\partial\phi}{\sqrt{V(\phi)}}.\label{change var}\ee As we will show quite soon, the above field re-definition allows to introduce normalized phase space variables that have been formerly used in similar studies with the application of the dynamical systems tools. In terms of these variables one can obtain a closed autonomous system of ODE out of the cosmological field equations written in terms of the transformed tachyon field $\phi$, for a broad class of self-interaction potentials $V(\phi)$.

In what follows we shall consider a simplified scenario, where the Einstein's field equations in (flat) FRW metric, are sourced by a mixture of a perfect barotropic fluid with energy density and pressure $\rho_m$ and $p_m=\omega_m\rho_m$ respectively ($\omega_m$ is the equation of state parameter of the perfect fluid which, for dust, vanishes), and of a tachyon field $\vphi$. The field equations that follow from (\ref{sen}) are:

\bea &&2\dot H=-\left(\omega_m+1\right)\rho_m-\frac{\dot\vphi^2 V(\vphi)}{\sqrt{1-\dot\vphi^2}},\nonumber\\
&&\frac{\ddot\vphi}{1-\dot\vphi^2}+3H\dot\vphi=-\frac{\partial_\vphi V}{V}.\label{tachy feqs}\eea 

Under the change of variable (\ref{change var}) above, the action (\ref{sen}) and the corresponding field equations (\ref{tachy feqs}) -- complemented with the continuity equation of the perfect fluid and the Friedmann equation -- are transformed into:

\bea &&S=\frac{1}{2}\int d^4x\sqrt{|g|}R+S_m+S_\phi,\nonumber\\&&S_\phi=-\int d^4x\sqrt{|g|}V(\phi)\sqrt{1+(\partial\phi)^2/V(\phi)},\label{tachyon action}\eea and

\bea &&3H^2=\rho_m+\rho_\phi,\nonumber\\
&&2\dot H=-\left(\omega_m+1\right)\rho_m-(\rho_\phi+p_\phi),\nonumber\\
&&\dot\rho_m=-3\left(\omega_m+1\right) H\rho_m,\nonumber\\
&&\ddot\phi+3\gamma^{-2}H\dot\phi=-\partial_\phi V(1-3\dot\phi^2/2V),\label{feqs}\eea respectively, where $$\rho_\phi=\gamma V,\;\;p_\phi=-\gamma^{-1}V,$$ and the Lorentz boost $\gamma$ is defined as:

\be\gamma=\frac{1}{\sqrt{1-\dot\phi^2/V}}.\label{modlorentzfactor}\ee 

In the next section we will study the dynamics of the transformed model described by equations (\ref{feqs},\ref{modlorentzfactor}) which leads to a closed autonomous system of ODE, as we shall see. The obtained results, if desired, can be translated in terms of field variables of the original tachyon model (\ref{sen}).

\section{Dynamical System}

Finding exact solutions of the cosmological equations (\ref{feqs}) is, in general, a very difficult task. That is why we will rely on the dynamical systems tools to investigate the asymptotic structure of the proposed tachyon cosmological model. To this end we will apply the concise recipes given in the appendix (section \ref{VII}). The goal will be to write the system of cosmological equations (\ref{feqs}) in the form of an autonomous system of ODE, as described in the appendix, so that one could associate such important dynamical systems concepts as past and future attractors (also saddle equilibrium points), with dynamical configurations -- solutions -- of the models. This is a powerful approach to uncover the most generic classes of solutions that are allowed by them. In order to build an autonomous system out of the system of cosmological equations (\ref{feqs},\ref{modlorentzfactor}) we introduce the following dimensionless phase space variables:

\be  x\equiv\frac{\dot\phi}{\sqrt{V}},\;y\equiv\frac{\sqrt{V}}{\sqrt{3}H}\;\Rightarrow\;\gamma=\frac{1}{\sqrt{1-x^2}}.\label{variables}\ee After this choice of phase space variables we can write the following autonomous system of ordinary differential equations:

\bea&&x'=(x^2-1)\{3x+\sqrt 3(\partial_\phi\ln
V)y\},\label{eqx}\\&&y'=\frac{y}{2}\left\{\sqrt{3}xy(\partial_\phi\ln V)+3\gamma_m+\right.\nonumber\\&&\;\;\;\;\;\;\;\;\;\;\;\;\;\;\;\;\;\;\;\;\;\;\;\;\;\;\;\;\left.+\frac{3y^2(x^2-\gamma_m)}{\sqrt{1-x^2}}\right\},\label{eqy}\eea where we have introduced the barotropic index for matter $\gamma_m=\omega_m+1$ (do not confound with $\gamma$ which plays the role of the Lorentz boost), and the prime denotes derivative with respect to the time variable $\tau\equiv\ln a$ -- properly the number of e-foldings of expansion. While deriving the ordinary differential equations (\ref{eqx}), and (\ref{eqy}), the following Friedmann constraint has also been considered:

\be
\Omega_m\equiv\frac{\rho_m}{3H^2}=1-\frac{y^2}{\sqrt{1-x^2}}.\label{constraint}\ee

It will be helpful to have other parameters of observational importance such as $\Omega_\phi=\rho_\phi/3H^2$ -- the scalar field dimensionless energy density parameter, and the equation of state (EOS) parameter $\omega_\phi=p_\phi/\rho_\phi$, written in terms of the variables of phase space:

\be \Omega_\phi=\frac{y^2}{\sqrt{1-x^2}},\;\;\omega_\phi=x^2-1.\ee Additionally, the deceleration parameter $q=-(1+H'/H)$: 

\be
q=-1+\frac{3}{2}\left[\gamma_m+\frac{y^2(x^2-\gamma_m)}{\sqrt{1-x^2}}\right].\ee

\subsection{Exponential Potential}\label{expo}

For an exponential self-interaction potential of the form: $$V(\phi)=V_0\exp(-\lambda\phi),$$ since $\partial_\phi\ln V=-\lambda=const$, then the equations (\ref{eqx}) and (\ref{eqy}) form a closed autonomous system of ODE:

\bea&&x'=(x^2-1)\{3x-\sqrt 3\lambda
y\},\nonumber\\&&y'=\frac{y}{2}\left\{3\gamma_m-\sqrt{3}\lambda
xy+\frac{3y^2(x^2-\gamma_m)}{\sqrt{1-x^2}}\right\}.\label{odeExpo}\eea

The phase space where to look for equilibrium points of the system of ODE (\ref{odeExpo}) -- corresponding to the model described by (\ref{feqs}) -- can be defined as follows:

\be \Psi=\{(x,y):-1\leq x\leq 1,\;0\leq y^4\leq 1-x^2\}.\label{psi}\ee

The equations (\ref{odeExpo}) coincide with Eqs. (8,9) of \cite{Copeland}, for the untransformed tachyon field with inverse square-law self-interaction potential of the form: $$V(\vphi)=V_0\vphi^{-2}.$$ Hence, the same critical points as in \cite{Copeland} are found in the present case, this time for the exponential potential.\footnote{Recall, however, that in the above reference the variable $x$ is defined in a simpler way: $x\equiv\dot\phi$.}

\begin{figure}[t!]
\begin{center}
\includegraphics[width=5.5cm,height=5cm]{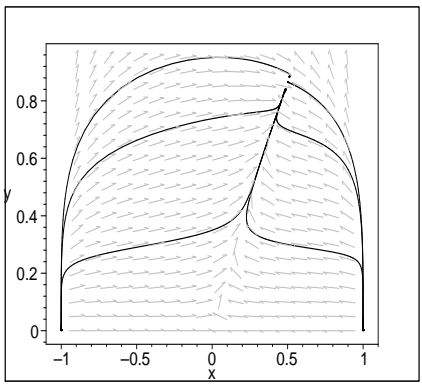}
\vspace{0.3cm}\bigskip\caption{Trajectories in phase space for given initial data, for the exponential potential. The free parameters have been arbitrarily set to $\gamma_m=0.25$ and $\lambda=1$. The matter-scaling solution is the late-time attractor, while the potential/kinetic energy scaling solution is the past attractor in the phase space.}
\label{fig1}
\end{center}
\end{figure}

In the Figure \ref{fig1} we show the trajectories in phase space for different sets of initial conditions for the model driven by an exponential potential. The free parameters have been arbitrarily set to $\gamma_m=0.25$ and $\lambda=1$. Due to this choice of the parameters, the equilibrium points $U=(\pm 1,0)$ represent inflationary critical points -- past attractors -- in the phase space, while the matter-scaling solution is the late-time attractor. Points $U$ are associated with ultra-relativistic behavior since $x\rightarrow\pm 1\;\Rightarrow\;\gamma\rightarrow\infty$. 

Due to our definition of the variable $x\equiv\dot\phi/\sqrt V$, the past attractor represents a scaling of the potential and of the kinetic energy of the tachyon scalar $$x=\pm 1\;\Rightarrow\;\dot\phi^2=V(\phi).$$ This means that, in case the inflationary past attractor be identified with early-time inflation, it can not be associated with the slow-roll approximation which implies that the potential energy of the scalar field dominates over the kinetic one (the latter may be disregarded).

In the next section we will focus on the asymptotic properties of models driven by several others self-interaction potentials of cosmological relevance.

\begin{table*}[tbp]\caption[crit]{Properties of the critical points of the autonomous system (\ref{ode}) for the inverse power-law potential $V(\phi)=V_0 \phi^{-\lambda}$. For points $M$ and $U$ the variable $v$ can take any value within the phase space, including $v=0$.}
\begin{tabular}{@{\hspace{4pt}}c@{\hspace{14pt}}c@{\hspace{14pt}}c@{\hspace{14pt}}c@{\hspace{14pt}}c@{\hspace{14pt}}c@{\hspace{14pt}}c@{\hspace{14pt}}c}
\hline\hline\\[-0.3cm]
Equilibrium Point &$x$&$y$&$v$&Existence& $\Omega_\phi$& $\omega_\phi$& $q$\\[0.1cm]
\hline\\[-0.2cm]
%%%%%%%%
$M$ & $0$ & $0$ & $v$ & Always & $0$ & $-1$ & $\frac{3\gamma_m-2}{2}$ \\[0.2cm]
$dS$ & $0$ & $1$ & $0$ & '' & $1$ & $-1$ & $-1$ \\[0.2cm]
$U$ & $\pm 1$ & $0$ & $v$ & '' & $0$ & $0$ & $\frac{3\gamma_m-2}{2}$ \\[0.2cm]
\hline \hline
\end{tabular}\label{tab1}
\end{table*}
\begin{table*}[tbp]\caption[eigenv]{Eigenvalues of the linearization matrices for the critical points in table \ref{tab1}.}
\begin{tabular}{@{\hspace{4pt}}c@{\hspace{14pt}}c@{\hspace{14pt}}c@{\hspace{14pt}}c@{\hspace{14pt}}c@{\hspace{14pt}}c@{\hspace{14pt}}c}
\hline\hline\\[-0.3cm]
Equilibrium Point&$x$&$y$& $v$& $\lambda_1$& $\lambda_2$& $\lambda_3$\\[0.1cm]\hline\\[-0.2cm]
%%%%%%%%
$M$ &$0$&$0$&$v$ & $-3$ &$0$& $3\gamma_m/2$\\[0.2cm]
$dS$ &$0$&$1$&$0$ & $-3$ &$0$& $-3\gamma_m$\\[0.2cm]
$U$ &$\pm 1$&$0$&$v$ & $6$ &$0$& $3\gamma_m/2$\\[0.2cm]
\hline \hline
\end{tabular}\label{tab2}
\end{table*}

\subsection{Self-interaction Potentials beyond the Exponential one}

As long as one considers just constant and exponential self-interaction potentials ($\partial_\phi\ln V=0$ and $\partial_\phi\ln V=const$ respectively), the equations (\ref{eqx}) and (\ref{eqy}) form a closed autonomous system of ODE. However, if one wants to go further to consider a wider class of self-interaction potentials beyond the exponential one, the system of ODE (\ref{eqx},\ref{eqy}) is not a closed system of equations any more, since, in general, $\partial_\phi\ln V$ is a function of the scalar field $\phi$. A way out of this difficulty can be based on the method developed in \cite{chinos}. In order to be able to study arbitrary self-interaction potentials one needs to consider one more variable $v$, that is related with the derivative of the self-interaction potential through the following expression

\be v\equiv-\partial_\phi V/V=-\partial_\phi\ln
V.\label{s}\ee Hence, an extra equation

\be v'=-\sqrt{3} x y v^2 (\Gamma-1),\label{sn'}\ee  has to be added to the above autonomous system of equations. The quantity $\Gamma\equiv V\partial_\phi^2 V/(\partial_\phi V)^2$ in equation (\ref{sn'}) is, in general, a function of $\phi$. The idea behind the method in \cite{chinos} is that $\Gamma$ can be written as a function of the variable $v$, and, perhaps, of several constant parameters. Indeed, for a wide class of potentials the above requirement -- $\Gamma=\Gamma(v)$ --, is fulfilled. Let us introduce a new function $g(v)=v^2(\Gamma(v)-1)$ so that equation (\ref{sn'}) can be written in the more compact form:

\be v'=-\sqrt{6} x y g(v).\label{snn'}\ee Equations (\ref{eqx}), (\ref{eqy}), and (\ref{snn'}) form a three-dimensional (closed) autonomous system of ODE:

\bea &&x'=(x^2-1)\{3x-\sqrt 3 y v\},\nonumber\\
&&y'=\frac{y}{2}\{3\gamma_m-\sqrt 3 x y v+\frac{3(x^2-\gamma_m)y^2}{\sqrt{1-x^2}}\},\nonumber\\
&&v'=-\sqrt{3} x y g(v),\label{ode}\eea that can be safely studied with the help of the standard dynamical systems tools. The function $g(v)$ can be analytically written for a wide variety of potentials. If $g(v)$ were a polynomial in $v$ (it is the case for most quintessential potentials of cosmological interest), then, with each root $v=v_{0i}$ of the polynomial equation $g(v)=0$ (the $v_{0i}$-s include the non-vanishing roots of the polynomial as well as the trivial root $v=0$), it can be associated an equilibrium point of the autonomous system of ODE (\ref{ode}). When $v=0$, $V(\phi)=V_0$, while, if $v_{0i}\neq 0$, then $V_i(\phi)\propto\exp{(-v_{0i}\phi)}$. Therefore, equilibrium points that arise under the requirement $g(v)=0$, are associated either with exponential potentials in the DBI-field $\phi$ ($v_{0i}\neq 0$), or with the constant potential ($v=0$). 

A rough inspection of the ODE (\ref{ode}) reveals that, independent on $g(v)$, for critical pints with $y=0$, then, either $x=\pm 1$, or $x=0$. Therefore, independent on the functional form of the potential $V(\phi)$, the kinetic/potential energy-scaling solution ($x=\pm 1\;\Rightarrow\;\dot\phi^2=V(\phi)$), and the matter-dominated solution ($x=0$, $y=0$, i. e., $3H^2=\rho_m$) are always equilibrium points of the autonomous system of ODE (\ref{ode}). Alternatively, for equilibrium points with $x=0$ ($y\neq 0$), $$x'=\sqrt 3 yv,\;y'=\frac{3\gamma_m}{2}y(1-y^2)\;,$$ so that, necessarily $y=\pm 1$, $v=0$. This means that for polynomials $g(v)$, for which $v=0$ is a root of the polynomial equation $g(v)=0$, the de Sitter solution $3H^2=V_0$ is an equilibrium point of the autonomous system (\ref{ode}).

If one started with the untransformed tachyon cosmological equations (\ref{tachy feqs}) instead, then, due to the square root $\sqrt{1-\dot\vphi^2}$, one were forced to choose the phase space variables \cite{Copeland} $$x\equiv\dot\vphi,\;\;y\equiv\frac{\sqrt{V}}{\sqrt 3 H},\;\;\lambda\equiv-\frac{\partial_\vphi V}{V^{3/2}}.$$ The latter variable is necessary to close the corresponding system of ODE (compare with our variable $v$). Contrary to the case with the variable $v$, it can be shown that it is very difficult to write the parameter $\Gamma\equiv V\partial_\vphi^2 V/(\partial_\vphi V)^2$ for arbitrary potentials, as a function of the variable $\lambda$. In fact, only power-law potentials, in particular $V(\vphi)\propto\vphi^{-2}$, can be analytically investigated \cite{Copeland}. To study the dynamics driven by other self-interaction potentials one has to rely on the obscure concept of "instantaneous critical points". 

At this point one recognizes the importance of the tachyon field transformation (\ref{change var}). Actually, as already shown, the cosmological equations of tachyon cosmology (\ref{tachy feqs}) are mathematically equivalent to (\ref{feqs}) under the change of variable (\ref{change var}). Therefore, one might work in terms of the transformed tachyon field $\phi$ (including interpretation of the results), or just to transform back the results of the dynamical systems study to the untransformed theory with the help of the inverse transformation $$\phi\rightarrow\vphi=\int\frac{d\phi}{\sqrt{V(\phi)}}\;.$$ In consequence, the present approach, which is based on the combined application of the transformation (\ref{change var}) of the field variable and of a method formerly used, for instance in Ref.\cite{chinos} (see above in this subsection), enables us to further generalize former studies -- with the application of the dynamical systems tools to tachyon cosmology -- to a wider variety of self-interaction potentials. 

A drawback of the approach undertaken here is inherited from the method of Ref.\cite{chinos}. It is associated with the fact that, for potentials that vanish at the minimum -- usually correlated with late-time dynamics -- the variable $v$ is undefined so that the corresponding behavior can not be properly investigated in the phase space spanned by $x,y,v$.

The phase space where to look for equilibrium points of the system of ODE (\ref{ode}), corresponding to the transformed model, can be defined as follows (we take into account only expanding universes so that only $y\geq 0$ are being considered):

\be \Psi=\{(x,y,v):-1\leq x\leq 1,\;0\leq y^4\leq
1-x^2,\;v\},\label{psi'}\ee where it has to be pointed out that the range of the variable $v$ depends on the specific kind of self-interaction potential considered. Recall that, for constant and exponential self-interaction potentials one does not need to consider the latter
variable, so that, the corresponding system of ODE is a two-dimensional one. 

In the next subsections we study particular examples were the usefulness of the approach undertaken here is illustrated.

\begin{figure}[t!]\begin{center}
\includegraphics[width=4cm,height=3.5cm]{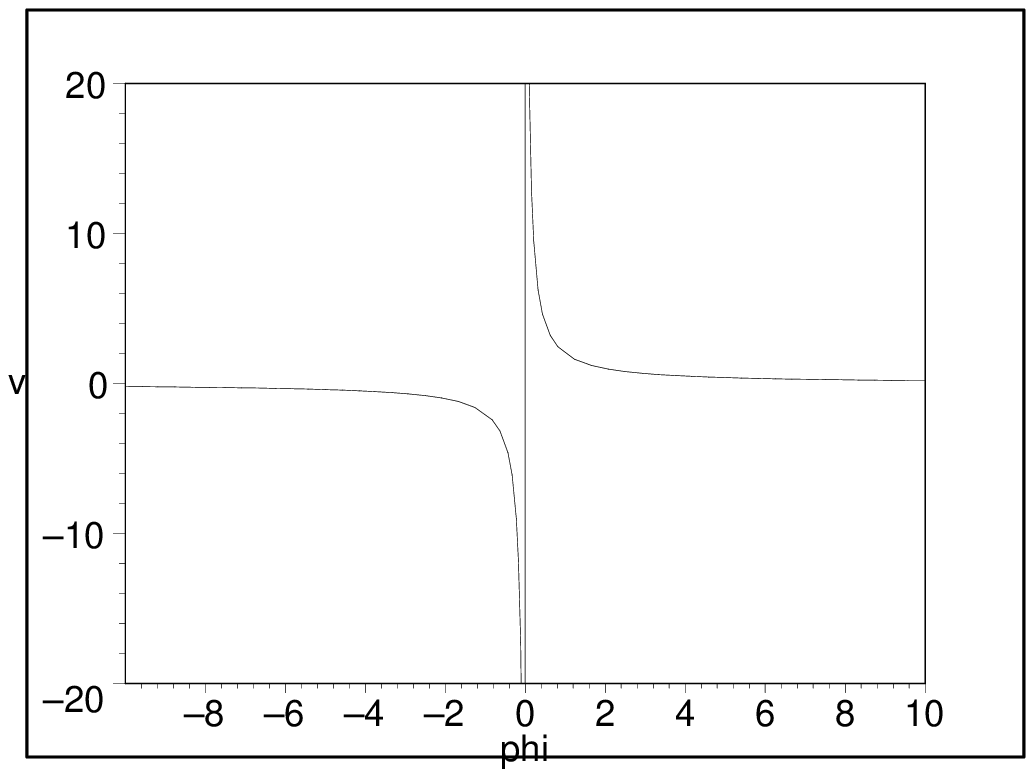}\includegraphics[width=4cm,height=3.5cm]{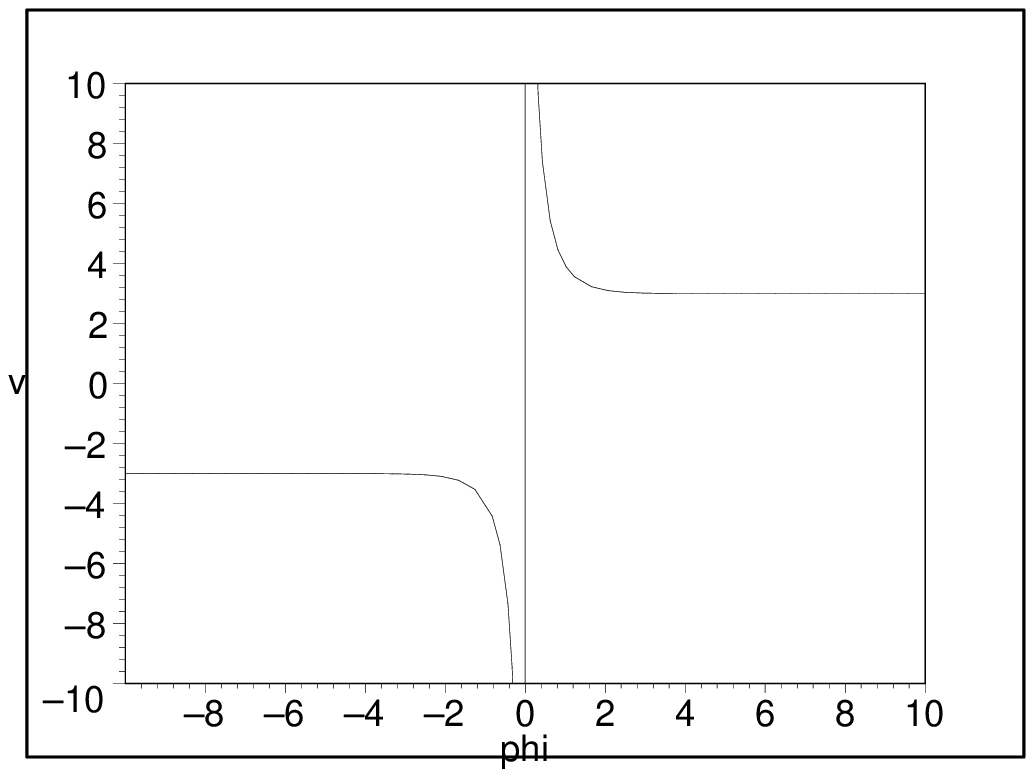}
\vspace{0.3cm}\bigskip\caption{A plot of the variable $v$ vs $\phi$ for the potential $V=V_0 \phi^{-\lambda}$ is shown for the chosen values of the free parameters: $V_0=1$, $\lambda=2$ (left-hand panel), while the same plot for the potential $V=V_0 \sinh^{-\alpha}(\lambda\phi)$ is shown in the right-hand panel for $V_0=1$, $\lambda=1$ and $\alpha=3$.}
\label{fig2}
\end{center}
\end{figure}

\begin{table*}[tbp]\caption[crit]{Properties of the critical points of the autonomous system (\ref{ode}) for the potential $V=V_0 [\sinh(\lambda\phi)]^{-\alpha}$.\\Here $v\geq\alpha\lambda$ (the constant potential $v=0$ corresponds to the particular case where either $\alpha=0$, or $\lambda=0$), and we introduced the following constant parameter: $y_*^2\equiv(\sqrt{36-\alpha^4\lambda^4}-\alpha^2\lambda^2)/6$.}
\begin{tabular}{@{\hspace{4pt}}c@{\hspace{14pt}}c@{\hspace{14pt}}c@{\hspace{14pt}}c@{\hspace{14pt}}c@{\hspace{14pt}}c@{\hspace{14pt}}c@{\hspace{14pt}}c}
\hline\hline\\[-0.3cm]
Eq. Point&$x$&$y$&$v$&Existence& $\Omega_\phi$& $\omega_\phi$& $q$\\[0.1cm]
\hline\\[-0.2cm]
%%%%%%%%
$M$ & $0$ & $0$ & $v$ & Always & $0$ & $-1$ & $(3\gamma_m-2)/2$ \\[0.2cm]
$U$ & $\pm 1$ & $0$ & $v$ & " & $0$ & $0$ & $(3\gamma_m-2)/2$ \\[0.2cm]
$T$ & $\alpha\lambda y_*/\sqrt{3}$ & $y_*$ & $\alpha\lambda$ & '' & $1$ & $-y_*^2/6$ & $(3\gamma_m-2-\alpha^2\lambda^2 y_*^2)/2$ \\[0.2cm]
$MS$ & $\sqrt{\gamma_m}$ & $\sqrt{3\gamma_m}/\alpha\lambda$ & $\alpha\lambda$ & $3\gamma_m\leq\alpha^2\lambda^2y_*^2$ & $3\gamma_m/\alpha^2\lambda^2\sqrt{1-\gamma_m}$ & $-1+\gamma_m$ & $(3\gamma_m-2)/2$ \\[0.2cm]
\hline \hline
\end{tabular}\label{tab3}
\end{table*}
\begin{table*}[tbp]\caption[eigenv]{Eigenvalues of the linearization matrices corresponding to the equilibrium points in Tab.\ref{tab3}.\\Here $\Pi\equiv\sqrt{(48\gamma_m^2\sqrt{1-\gamma_m}/\alpha^2\lambda^2)+4+\gamma_m(17\gamma_m-20)}$.}
\begin{tabular}{@{\hspace{4pt}}c@{\hspace{14pt}}c@{\hspace{14pt}}c@{\hspace{14pt}}c@{\hspace{14pt}}c@{\hspace{14pt}}c@{\hspace{14pt}}c}
\hline\hline\\[-0.3cm]
Eq. Point&$x$&$y$& $v$& $\lambda_1$& $\lambda_2$& $\lambda_3$\\[0.1cm]\hline\\[-0.2cm]
%%%%%%%%
$M$ &$0$&$0$&$v$ & $-3$ &$0$& $3\gamma_m/2$\\[0.2cm]
$U$ &$\pm 1$&$0$&$v$ & $6$ &$0$& $3\gamma_m/2$\\[0.2cm]
$T$ &$\alpha\lambda y_*/\sqrt{3}$&$y_*$&$\alpha\lambda$& $-2\alpha\lambda^2y_*$ &$-3+\alpha^2\lambda^2y_*^2/2$& $-3\gamma_m+\alpha^2\lambda^2y_*^2$\\[0.2cm]
$MS$ &$\sqrt{\gamma_m}$&$\sqrt{3\gamma_m}/\alpha\lambda$&$\alpha\lambda$ & $-6\gamma_m/\alpha$ &$-3[(2-\gamma_m)+\Pi]/4$& $-3[(2-\gamma_m)-\Pi]/4$\\[0.2cm]
\hline \hline
\end{tabular}\label{tab4}
\end{table*}

\subsubsection{The (inverse) power-law potential $V(\phi)=V_0\phi^{-\lambda}$}\label{powerlaw}

The inverse power-law potential have been extensively studied within untransformed tachyon field model \cite{Copeland,finelli}. According to the definition (\ref{s}) of the variable $v$, for this potential one has: 

\be v=\lambda \phi^{-1},\label{vpot1}\ee so that the following asymptotics hold true:

\be \lim_{\phi\rightarrow\pm 0}v(\phi)=\pm\infty\;,\;\;\;\lim_{\phi\rightarrow
\pm\infty}v(\phi)=0.\ee

In the left hand panel of Figure \ref{fig2} a plot of $v(\phi)$ vs $\phi$ is shown for the chosen values of free parameters: $V_0=1$, $\lambda=2$. Notice that the range $\phi\in ]-\infty,0[$ is covered by negative values of the variable $v$, while positive values of $v>0$ cover the range $\phi\in ]0,\infty[$. In what follows, for definiteness, we will restrict ourselves to $v>0$, so that the tachyon field variable takes values in the interval $\phi\in]0,\infty[$. In this case the function $g(v)$ in (\ref{snn'},\ref{ode}) can be written in the following way: 

\be g(v)=v^2/\lambda.\label{fv1}\ee 

The cosmic dynamics driven by this potential can be associated with a 3-dimensional phase space (\ref{psi'}), spanned by the
variables $x$, $y$, and $v$, where $0\leq v<\infty$.

The equilibrium points of the autonomous system of ODE (\ref{ode}) in the phase space $\Psi$ defined above, are listed in table \ref{tab1} while the eigenvalues of the corresponding linearization matrices are shown in table \ref{tab2}. These are non-hyperbolic critical points meaning that only limited information on their stability properties can be retrieved by means of the present linearized analysis. To help us understanding the stability properties of these points one has to rely on the phase portraits which show the structure of the phase space through phase path-probes originated by given initial data. 

Existence of the matter-dominated solution (equilibrium point $M$ in table \ref{tab1}), is independent on the value of the variable $v$, meaning that this phase of the cosmic evolution may arise at early-to-intermediate times ($0<v<\infty$), as well as at late
time ($v=0$). As seen from table \ref{tab2}, since in this case the two non vanishing eigenvalues of the linearization matrix are of opposite sign, the matter-dominated solution is always a saddle equilibrium point of (\ref{ode}). This solution is inflationary whenever $\gamma_m<2/3$. 

Something similar can be said about the equilibrium point $U$ in Tab.\ref{tab1}. This point can be associated also either with early-time, as well as with intermediate-time dynamics, due to the fact that $v$ can be any value. However, unlike the matter-dominated solution, the critical point $U$ could be a past attractor (source critical point, usually associated with early-time dynamics) in the phase space, since the two non-vanishing eigenvalues of the corresponding linearization matrix are positive. Notwithstanding, since the point is a non-hyperbolic one, no final statement can be made on its stability properties until the corresponding phase portraits are drawn. The point $U$ is also a matter-dominated solution, however, this phase corresponds to the ultra-relativistic case (the Lorentz boost $\gamma\rightarrow\infty$) and, unlike the standard situation, it represents scaling between the kinetic and the potential energy densities of the tachyon scalar. As before, the solution is inflationary whenever $\gamma_m<2/3$.

The late-time dynamics driven by the potential $V=V_0 \phi^{-\lambda}$ is correlated with infinitely large values of the variable $\phi\rightarrow\infty$ which, in the phase space (\ref{psi'}) ($0\leq v<\infty$), is depicted by the equilibrium point with $v=0$ in table \ref{tab1} (equilibrium point $dS$). This point corresponds to the inflationary de Sitter solution $(3H^2=V)$. From Tab.\ref{tab2} it is seen that this equilibrium point could be a late-time attractor since the two non-vanishing eigenvalues of the corresponding linearization matrix are both negative. However, since as already said, this is a non-hyperbolic point, only after drawing the corresponding phase portraits one is able to make conclusive statements about its stability properties. 

We want to notice that the above results remain true if one considered power-law potentials with negative values of the constant parameter $\lambda$. In particular the above results can be safely extended to the quadratic potential $V(\phi)\propto\phi^2$.

\begin{figure}[t!]
\begin{center}\includegraphics[width=4cm,height=3.5cm]{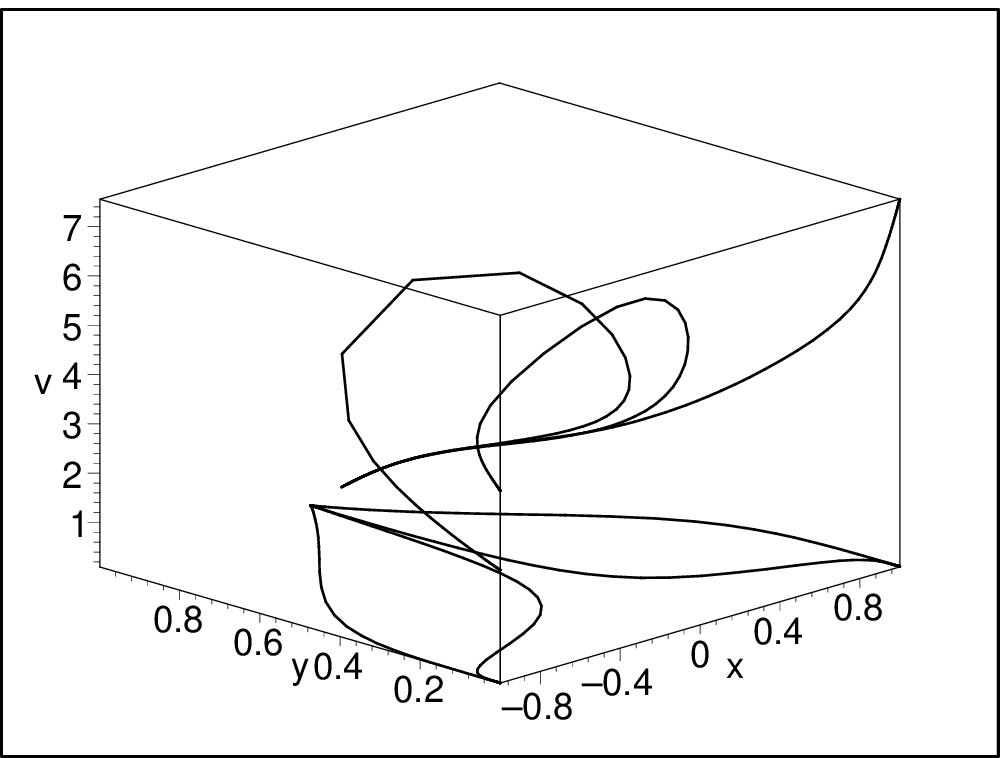}\includegraphics[width=4cm,height=3.5cm]{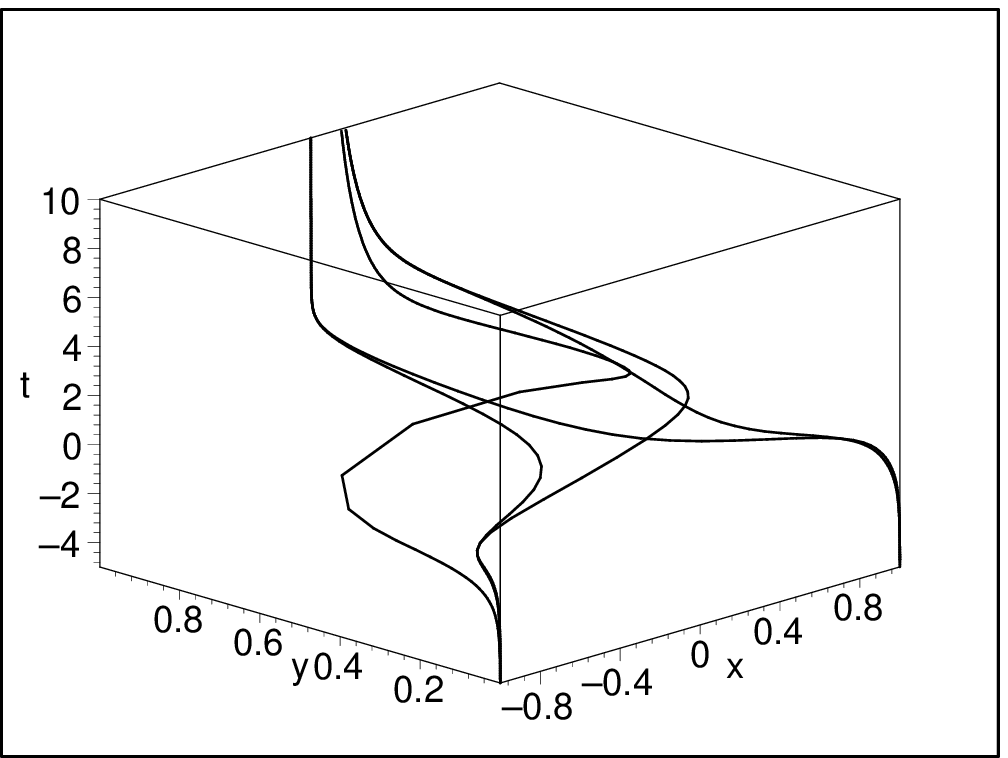}
\vspace{0.3cm}\bigskip\caption{In the left-hand panel trajectories in phase space for different sets initial conditions are drawn for the potential $V=V_0 \phi^{-\lambda}$ ($V_0=1$, $\lambda=2$, $\gamma_m=1$ - background dust), while the flux in time of the corresponding system of ODE is depicted in the right-hand panel of the figure. Probe-paths in the phase space originate at the ultra-relativistic equilibrium point $U$ ($x=\pm 1$) while at late times approach to the de Sitter attractor $dS$ ($x=0,\;y=1$).}\label{fig3}
\end{center}
\end{figure}

In Fig.\ref{fig3}, in order to illustrate the stability properties of the asymptotic solutions $M$, $dS$, and $U$, probe paths in phase space -- trajectories in the phase space originated by given initial data -- are drawn. As clearly seen, these trajectories emerge from the ultra-relativistic (matter-dominated) point $U$ and converge towards the inflationary de Sitter attractor (point $dS$) at late times, thus confirming our suspects that $U$ is the past attractor, while $dS$ is the future attractor. 

\subsubsection{The potential $V(\phi)=V_0\left[\sinh(\lambda\phi)\right]^{-\alpha}$.}\label{sinh}

This potential has been formerly studied in Ref.\cite{sahni1} as a new cosmological tracker solution for quintessence. According to the definition (\ref{s}), for this potential one gets: 

\be v=\alpha\lambda \coth(\lambda\phi),\label{vpot2}\ee from which it follows, in particular, that 

\be \lim_{\phi\rightarrow 0}v(\phi)=\infty\;,\;\;\;\lim_{\phi\rightarrow
\pm\infty}v(\phi)=\pm\alpha\lambda.\ee

In the right-hand panel of Fig.\ref{fig2} a plot of $v(\phi)$ vs $\phi$ is shown for the chosen values of free parameters: $V_0=1$, $\lambda=1$ and $\alpha=3$. Notice that the range of the variable $\phi\in ]-\infty,0[$ is covered by $-\infty<v\leq -\alpha\lambda$, while the range $\phi\in ]0,\infty[$ is covered by $\alpha\lambda\leq v<\infty$. In what follows, for definiteness we will restrict ourselves to the interval $\alpha\lambda\leq v<\infty$. For the above potential the function $g(v)$ defined in (\ref{snn'},\ref{ode}) can be written in the following way: 

\be g(v)=\frac{v^2-\alpha^2\lambda^2}{\alpha}.\label{fv2}\ee 

The cosmic dynamics driven by $V=V_0 [\sinh(\lambda\phi)]^{-\alpha}$ can be associated with the 3-dimensional phase space (\ref{psi'}), where the variable $v$ is constrained to the interval $\alpha\lambda\leq v<\infty$. The equilibrium points of the autonomous system of ODE (\ref{ode}) in the phase space $\Psi$ defined above, are listed in table \ref{tab3}, while the eigenvalues of the corresponding linearization matrices are shown in Tab.\ref{tab4}.

As for the power-law potential, the existence of the matter-dominated solution (equilibrium point $M$ in table \ref{tab3}), is independent of the value of the variable $v$, meaning that this phase of the cosmic evolution may arise at early-to-intermediate times ($\alpha\lambda<v<\infty$), as well as at late times ($v=\alpha\lambda$). As seen from Tab.\ref{tab4}, since in this case the two non vanishing eigenvalues of the linearization matrix are of opposite sign, the matter-dominated solution is always a saddle equilibrium point of (\ref{ode}). The corresponding cosmological solution represents decelerating expansion whenever $\gamma_m>2/3$. Unlike this, the matter-dominated equilibrium point $U$ in Tab.\ref{tab3} can be associated with ultra-relativistic behavior (large Lorentz boost). As already said this point represents scaling between the potential and the kinetic energies of the tachyon field. As it can be seen from the phase portraits, it is always the past attractor for any path in the phase space of the model.

Equilibrium points $T$ (the tachyon-dominated solution) and $MS$ (the matter-scaling solution) are associated with late-time dynamics since, according to (\ref{vpot2}), $v=\alpha\lambda$ is correlated with infinitely large values of the variable $\phi$. The scalar field-dominated solution $T$ always exits and whenever $\gamma_m>\alpha^2\lambda^2(\sqrt{36+\alpha^4\lambda^4}-\alpha^2\lambda^2)/18$ it is a stable equilibrium point (the late-time attractor), otherwise it is a saddle critical point in phase space. Whenever the matter-scaling solution $MS$ exists, it is a stable equilibrium point (the late-time attractor). This solution is always associated with accelerated expansion. As in Ref.\cite{Copeland}, one has to take caution since the critical point $MS$ does not exist if either $\gamma_m=1$ (matter-dominated era), or $\gamma_m=4/3$ (radiation domination). This is due to the fact that the existence of the matter-scaling solution requires fulfillment of the condition $0<\gamma_m< 1$. In this sense this solution can not be associated with a realistic model of dark energy. 

\begin{figure}[t!]
\begin{center}\includegraphics[width=4cm,height=4cm]{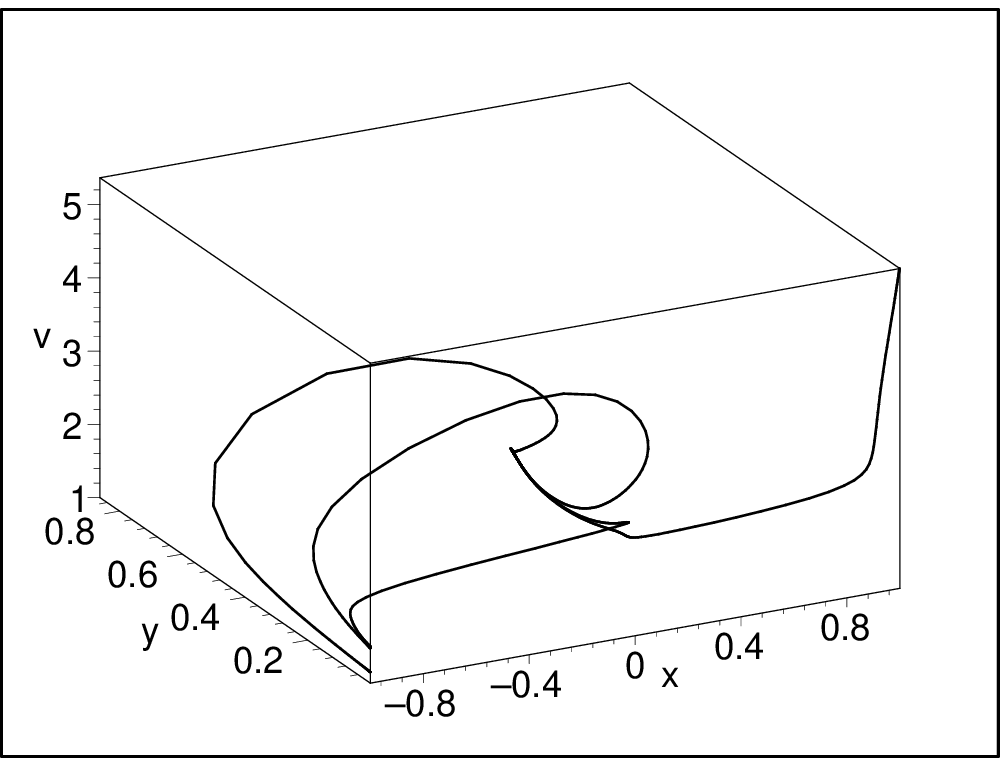}\includegraphics[width=4cm,height=4cm]{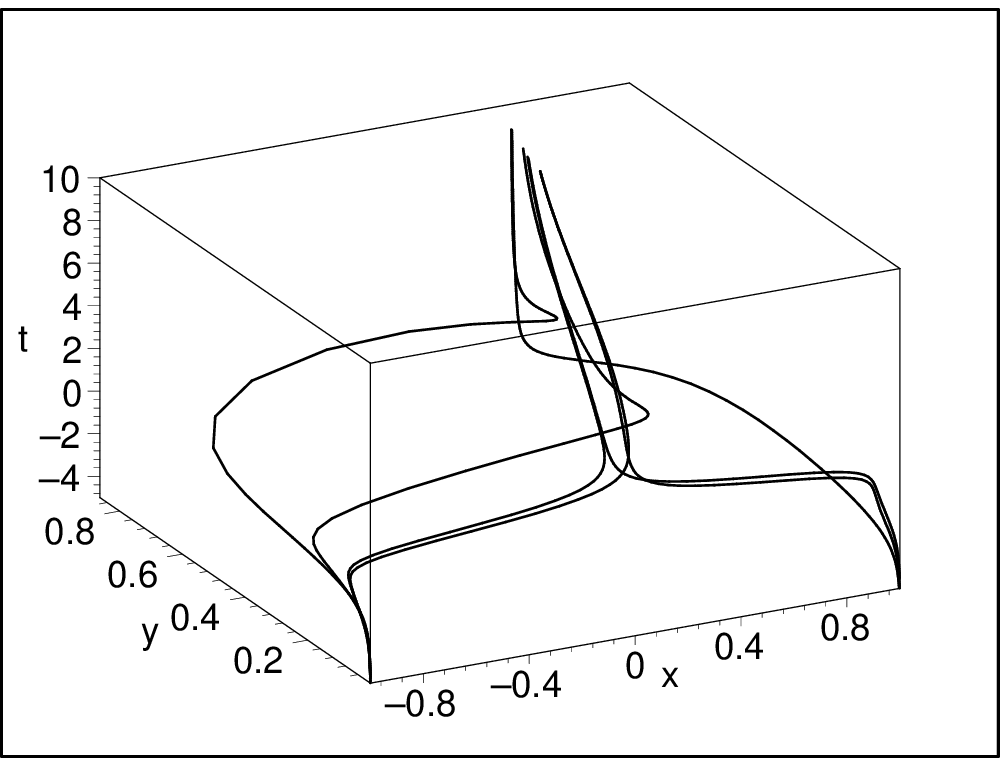}
\vspace{0.3cm}\bigskip\caption{Phase portrait for the model driven by the potential $V=V_0 [\sinh(\lambda\phi)]^{-\alpha}$ (left-hand panel), and the flux in time of the system of ODE (\ref{ode}) for this case (right-hand panel). The free parameters chosen are: $\gamma_m=.2$, $\lambda=1$ and $\alpha=1$. It is clearly seen that the ultra-relativistic solution $U$ is the past attractor while, due to the above choice of parameters, the matter-scaling solution $MS$ is the late-time attractor.}\label{fig4}
\end{center}
\end{figure}

In the figure \ref{fig4} the phase portrait for this case is depicted. The above discussed behavior is clearly illustrated by the figure. The free parameters were taken in such a way that the $MS$ solution exists ($\gamma_m=.2$, $\lambda=1$, $\alpha=1$). In correspondence, the ultra-relativistic phase $U$ is the past attractor, while the matter-scaling solution $MS$ is the late-time (inflationary) attractor.

\section{Correspondence between the transformed and the untransformed models}

From the point of view of their physical interpretation both models (\ref{sen}) and (\ref{tachyon action}) are equivalent. Actually, formal mathematical equivalence of both pictures under (\ref{change var}) implies a residual equivalence between magnitudes of physical relevance, in particular:

\be \Omega_\phi=\Omega_\vphi,\;\;\omega_\phi=\omega_\vphi,\;\;q(\phi)=q(\vphi)\equiv-(1+\frac{\dot H}{H^2}).\label{physical equivalence}\ee 

Anyway, as already mentioned, one might desire to have the results of the dynamical systems study in terms of the untransformed field variable $\vphi$. In such a case one has to perform the inverse transformation: $$\phi\rightarrow\vphi=\int\frac{d\phi}{\sqrt{V(\phi)}}.$$ It is evident that, once the functional form of the self-interaction potential $V(\phi)$ (or $V(\vphi)$) is known, the functional relationship $\vphi=\vphi(\phi)$ (or $\phi=\phi(\vphi)$) can be obtained through integration in quadratures, so that one is able to transform the potential $V(\phi)=V(\phi(\vphi))\;\rightarrow\;V=V(\vphi)$ (or $V(\vphi)=V(\vphi(\phi))\;\rightarrow\;V=V(\phi)$). Actually, from (\ref{change var}) it follows that

\be \vphi-\vphi_0=\int\frac{d\phi}{\sqrt{V(\phi)}},\;\text{or},\;\phi-\phi_0=\int\sqrt{V(\vphi)}d\vphi.\label{quad}\ee 

By using equation (\ref{quad}) it can be found that the transformation (\ref{change var}) implies the following correspondence between the transformed and untransformed tachyon potentials $V(\phi)$ and $V(\vphi)$:

\be V(\phi)=V_0 e^{-\lambda\phi}\;\rightarrow\;V(\vphi)=\bar V_0 (\vphi-\vphi_0)^{-2},\label{expo pot}\ee where $\bar V_0\equiv 4/(V_0\lambda^2)$, and $\vphi_0$ is an integration constant. For the inverse power-law potential one gets that

\be V(\phi)=V_0 \phi^{-2\lambda}\;\rightarrow\;V(\vphi)=\bar V_0 (\vphi-\vphi_0)^{-2n},\label{power pot}\ee where $\bar V_0=[V_0/(\lambda+1)^{2\lambda}]^{1/\lambda+1}$, and $n=\lambda/(\lambda+1)$. Additionally, for the sinh-like potential $\propto\sinh^{-2}(\lambda\phi)$, one obtains the following equivalence

\be V(\phi)=V_0 \sinh^{-2}(\lambda\phi)\;\rightarrow\;V(\vphi)=\frac{\bar V_0}{\vphi^2-\vphi_0^2},\label{sinh pot}\ee where $\bar V_0=1/\lambda^2$ and $\vphi_0^2=1/V_0\lambda^2$. 

Notice that under the transformation (\ref{change var}) the exponential potential is equivalent to the square-law tachyon potential $\propto\vphi^{-2}$ studied in \cite{Copeland}, while the (inverse) power-law potential $\propto\phi^{-2\lambda}$ is equivalent to a (inverse) power-law tachyon potential $\propto\vphi^{-2n}$, not fully investigated in the same reference. I. e., the inverse power-law potential is not transformed under (\ref{change var}). In the later case, the only difference of physical significance is in the power of the potential since $\lambda\rightarrow n=\lambda/(\lambda+1)$. It is seen that, for positive $\lambda>0\;\Rightarrow\;n\leq 1$.

The present approach can be applied to investigate the cosmic dynamics of a tachyon field for self-interaction potentials beyond the square-law potential which has been studied in detail in Ref.\cite{Copeland}. Actually, consider, for instance, the exponential self-interaction potential: $$V(\vphi)=V_0 e^{\mu\vphi}.$$ By using the relationship (\ref{quad}) it can be shown that: $$V(\vphi)=V_0 e^{\mu\vphi}\;\rightarrow\;V(\phi)=\frac{\mu^2}{4}\phi^2.$$ Fortunately this case has been already studied in subsection \ref{powerlaw} (just replace $\lambda\rightarrow -2$ in equations (\ref{vpot1}) and (\ref{fv1}), and bear in mind that the non-negative range of the variable $v$ is associated with negative values of the tachyon scalar $\phi\in]-\infty,0[$). It remains just to translate the corresponding results so that one could discuss their physical implications for the untransformed tachyon cosmological dynamics.

\section{Discussion}

Thanks to the formal mathematical equivalence between tachyon dynamics depicted by Sen's effective action (\ref{sen}) and the dynamics of the transformed field, given by the action (\ref{tachyon action}), under the transformation (\ref{change var}), the approach undertaken in this paper enables applying the standard tools of the dynamical systems to investigate the cosmic dynamics driven by a wide variety of self-interaction potentials, without resorting to such obscure concepts as ``instantaneous critical points'', whose physical relevance is suspicious. Actually, if such a mathematical (and dynamical) equivalence is taken into consideration, the results obtained in section III -- after applying the linear analysis to study the dynamics of the model of (\ref{tachyon action}) -- can be safely translated to the case of the tachyon model portrayed by (\ref{sen}). 

As shown in the former section there is a full correspondence between inverse power-law potential $\propto\phi^{-2\lambda}$ and that of the untransformed tachyon $\propto\vphi^{-2n}$, so that, for this kind of potential the results displayed in the tables \ref{tab1} and \ref{tab2} for the transformed $\phi$-field hold true for the tachyon cosmological model of \cite{Copeland}, which means in turn, that a detailed study of the dynamics driven by this tachyon potential is indeed possible. According to the results of section III (see Tabs. \ref{tab1} and \ref{tab2}), for the potential $V(\vphi)\propto\vphi^{-2n}$, whenever $0<n\leq 1$, one obtains that the de Sitter solution -- point $dS$ in Tab. \ref{tab1} -- is always the late-time attractor in the phase space, while the ultra-relativistic matter-dominated solution -- point $U$ in Tab. \ref{tab2} -- is the past attractor from which the phase paths originate. The matter-dominated solution $M$ is always a saddle in the phase space. Therefore, the standard tachyon cosmology model driven by the inverse power-law potential could be a nice scenario to address the late-time cosmic acceleration. 

The study of the asymptotic properties of the tachyon model for the exponential potential, $V(\vphi)\propto\exp(\mu\vphi)$, is mathematically equivalent to the study of the asymptotic properties of the transformed cosmological model for the quadratic potential $V(\phi)\propto\phi^2$, which is a particular case of the study presented in \ref{powerlaw} when the constant parameter $\lambda$ is replaced by the particular negative value $-2$. The only think to be changed is the phase space itself if one keeps $\phi\in]0,\infty[$ --- in this case, in place of the half of the phase space (\ref{psi'}) corresponding to positive $v$-s, one has to consider the complementary half defined by negative $v$-s instead ---, or one might keep intact the phase space at the cost that the tachyon field itself takes values in the interval $\phi\in]-\infty,0[$.

A remarkable property of the tachyon model studied here is that, independent of the particular functional form of the self-interaction potential $V(\phi)$ considered, the matter-dominated solutions $M$ and $U$ -- the ultra-relativistic matter-dominated solution, are always equilibrium points of the corresponding autonomous system of ODE (\ref{ode}) (see tables \ref{tab1},\ref{tab3}). A straightforward inspection of the equations (\ref{ode}) reveals why this happens. Actually, a crude inspection of the equations in the system of ODE (\ref{ode}) shows that, independent of the functional form of the function $g(v)$ and of the value of the variable $v$, since for $y=0$ the system (\ref{ode}) reduces to the simplified system of equations: $$x'=3x(x^2-1),\;\;\;y'=0,\;\;\;v'=0,$$ then, for $x=0$ and $x=\pm 1$, the corresponding points $(x,y,v)$ in phase space: $M=(0,0,v)$, and $U=(\pm 1,0,v)$, both are equilibrium points of the system of ODE (\ref{ode}). Since the existence of these points is independent of the value of the variable $v$, both phases of the cosmic evolution may arise at early, intermediate, as well as at late times. In fact, the point $M$ is always a saddle critical point, while $U$ is the past attractor for any path in the phase space of the model, otherwise, $U$ is the point in phase space from which all of the phase trajectories are repelled. 

From the analysis of the equations (\ref{ode}) it also arises that, in general, for potentials for which the function $g(v)$ is a polynomial in $v$ (it happens for most part of known quintessential potentials), and the polynomial equation $g(v)=0$ has non-vanishing roots $v=v_{0i}\neq 0$, since in this case the system (\ref{ode}) reduces to the autonomous system of ODE (\ref{odeExpo}) for an exponential potential $V\propto e^{\lambda\phi}$ ($\lambda=v_0$), the late-time dynamics of the tachyon field can be either the scalar field-dominated solution, or the matter-scaling phase. This conclusion is quite robust and has been formerly stated in \cite{chinos} in a different context.

\subsection{DBI/Tachyon equivalence}

We can use the field re-definition (\ref{change var}), to establish links between previous results within the DBI bibliography and the study of equivalent tachyon field cases. Actually, under (\ref{change var}) the tachyon action $S_\vphi$ in (\ref{sen}) is replaced by $S_\phi$ in (\ref{tachyon action}), which is a particular case of the generalized DBI-action \cite{speedlimit,basic,DBInflation}:\footnote{As it is the case for Sen's tachyon field, these generalized DBI-theories have attracted much attention in recent years due to their role in inflation based on string theory \cite{dvali}. In the above scenarios the inflaton is identified with the position of a mobile D3-brane, moving on a compact 6-dimensional submanifold of spacetime (for reviews and references see \cite{string}), which means that the inflaton is interpreted as an open string mode.}

\bea &&S_{DBI}=-\int d^4x\sqrt{|g|}\{f^{-1}(\phi)\sqrt{1+f(\phi)(\nabla\phi)^{2}}\nonumber\\
&&\;\;\;\;\;\;\;\;\;\;\;\;\;\;\;\;\;\;\;\;-f^{-1}(\phi)+V(\phi)\},\label{dbi action}\eea where $f(\phi)$ is the inverse of the brane tension (also acknowledged as the warp factor of the warped throat geometry in the internal space) and, $V(\phi )$ is the potential for the DBI-field, arising from quantum interactions between the D3-brane associated with $\phi$, and other D-branes. In the bibliography one usually encounters given forms of the warp factor $f(\phi)$. For instance, $f(\phi)=\lambda\phi^{-4}$ ($\lambda$-constant), or, also $f(\phi)=const.$ As one can immediately see, this action contains $S_\phi$ in (\ref{tachyon action}) as a particular case, if one chooses $f(\phi)$ in (\ref{dbi action}) to be related with the potential of the DBI-field in the following particular form: $f(\phi)\cdot V(\phi)=1$. 

In reference \cite{copeland-shuntaru}, for instance, the authors study (among other cases) the late-time dynamics of a DBI model with exponential potential and brane tension (here we use the notation of \cite{copeland-shuntaru}): $$V(\phi)=\sigma e^{-\lambda\phi},\;\;f(\phi)=\gamma e^{-\mu\phi}\;.$$ In the particular case where $\lambda+\mu=0\;\Rightarrow\;f\cdot V=\gamma\sigma$, a new class of solutions arise, for which $0<\tilde\gamma<1$ ($\tilde\gamma=\gamma^{-1}=\sqrt{1-x^2}$), i. e., $x\neq 0$, and $x\neq\pm 1$. These solutions are the matter-scaling solution and the scalar field-dominated (kinetic/potential energy-scaling) phase. It is interesting noting that, under the field replacement (\ref{change var}), and $f\cdot V=1$, the exponential potential above transforms into the (inverse) square tachyon potential $$V(\vphi)=\frac{4/\lambda^2}{(\vphi-\vphi_0)^2}\;,$$ so that, given that $\gamma\sigma=1$, the former results are equivalent to the results of the dynamical systems study of tachyon cosmology with an inverse square potential of reference \cite{Copeland}. Actually, the equilibrium points (c) and (d1,d2) in Tab. I of Ref. \cite{Copeland} are the above mentioned matter-scaling, and the scalar (tachyon) field-dominated solutions respectively. As long as we know, no such equivalence between generalized-DBI and tachyon dynamics has been reported before.

\section{Conclusions}

In the quest for alternative models of inflation (here we include both primordial and late-time inflation), the effective tachyon models and their generalization: the scalar DBI-field models, have played an important role. Due to complexity of the analysis, in particular in connection with difficulties to obtain closed (autonomous) systems of ordinary differential equations out of the corresponding cosmological equations, the study of the tachyon dynamics has been performed for a limited number of particular tachyon potentials.

In the present paper we have proposed an approach based on a re-definition of the tachyon field variable $\vphi$, followed by straightforward application of a method already used in the bibliography (see, for instance, \cite{chinos}) to study a wide class of self-interaction potentials in a different context. The above mentioned field-variable re-definition is necessary if one wants to introduce formerly used phase space variables that make the autonomous system of ODE -- obtained out of the corresponding cosmological equations -- a closed system of ODE. The present approach allows to apply the standard dynamical systems tools to the study of a broad class of self-interaction potentials, provided that the function $$g(v)\equiv v^2(\Gamma-1)\;,$$ where $$v\equiv-\frac{\partial_\phi V}{V},\;\;\Gamma=\frac{V\partial^2_\phi V}{(\partial_\phi V)^2}\;,$$ can be written as a polynomial in the new phase space variable $v$. The usefulness of our combined approach has been illustrated with the detailed study of the following tachyon potentials $$V(\vphi)=V_0\;\vphi^{-2n},\;\;V(\vphi)=\frac{V_0}{\vphi^2-\vphi^2_0}\;,$$ where the latter potential is just a particular case of the former one for $n=1$, $\vphi_0=0$. It has been demonstrated that the $\vphi^{-2n}$ potential is equivalent to the inverse power-law potential $\phi^{-2\lambda}$ ($n=\lambda/(\lambda+1)$), while the inverse square tachyon potential $\vphi^{-2}$ is equivalent to the exponential potential $\exp{-\lambda\phi}$. The particular inverse square tachyon potential $(\vphi^2-\vphi_0^2)^{-1}$ is equivalent to the sinh-like potential $\sinh^{-2}{(\lambda\phi)}$, therefore, different kinds of potentials in the transformed model might be equivalent to a same tachyon potential. In spite of the fact that only the dynamics driven by the inverse power-law tachyon potential $\propto\vphi^{-2n}$, and by the exponential one $\propto\exp(\vphi)$, have been studied in details in this paper, nonetheless, the combined method used here can be applied to other kinds of potentials, provided that the function $g(v)$ can be written as a polynomial in $v$ for the potential $V(\phi)$ of the transformed tachyon field $\phi$. 

An existing mathematical equivalence between the dynamics of a generalized DBI-field $\phi$ with the following relationship between the warp factor $f$ of the internal geometry, and the DBI potential $V$: $f\cdot V=1$, and the dynamics of a tachyon field $\vphi$ driven by the effective action $$S_\vphi=-\int d^4x\sqrt{|g|}V\sqrt{1+(\partial\vphi)^2}\;,$$ under the field redefinition $\vphi=\int d\phi/\sqrt{V(\phi)}$, has been also revealed. The above equivalence allows to translate the results of dynamical systems studies of DBI models with the condition $f\cdot V=1$, for given DBI potentials, to the corresponding studies of the equivalent tachyon models. 

Amongst the most interesting results of the present study we can list the following:

\begin{itemize}

\item It has been revealed that independent of the functional form of the potential, the matter-dominated solution and the ultra-relativistic (also matter-dominated) solution, are always associated with equilibrium points in the phase space of the tachyon models. The latter is always the past attractor, while the former is a saddle critical point.

\item We have demonstrated that, in general, for DBI potentials for which the function $g(v)$ can be written as a polynomial in $v$ (it is the case for most quintessential potentials of cosmological interest): $$g(v)=\sum_n (v-v_0)^{n}\;,$$ the late-time dynamics is associated with either the de Sitter solution if at the critical point $v=0$, is a root of the polynomial equation $g(v)=0$, or with matter-scaling and/or scalar field-dominated solutions if at the critical point, the given root $v=v_0\neq 0$. 

\item A quite trivial (formal) mathematical equivalence between DBI models with the following particular relationship: $f\cdot V=1$, between the warp factor $f$ and the DBI potential $V$, and tachyon cosmological models originated from the Lagrangian (\ref{sen}), under the field redefinition (\ref{change var}), allows to reveal that results of different, otherwise unrelated investigations, are fully equivalent. As an example: several results of studies of DBI models with exponential brane tension and potential in Ref.\cite{copeland-shuntaru}, and the results of the dynamical systems studies of tachyon cosmology with inverse square potential in reference \cite{Copeland}, have been shown to be equivalent. 

\end{itemize} 

The present approach can be safely (and moderately easy) applied to consider new classes of tachyon potentials beyond the (inverse) power-law and exponential ones. While the present paper was under consideration for publication, an alternative method to investigate the cosmic dynamics of a tachyon field for wide classes of potential, without performing any field re-definition, has been published \cite{last hour}.

This work was partly supported by CONACyT M\'{e}xico, under grants 49865-F, 54576-F, 56159-F, 49924-J, 105079, 52327, and by grant number I0101/131/07 C-234/07, Instituto Avanzado de Cosmologia (IAC) collaboration. T G, D G and Y N acknowledge the MES of Cuba for partial support of the research. R G-S acknowledges partial support from COFAA-IPN and EDI-IPN grants, and SIP-IPN 201006010.

\section{Appendix: Dynamical Systems}\label{VII}

Here we include brief tips of how to apply the dynamical systems tools to situations of cosmological interest. In order to apply these tools one has to follow the steps enumerated below: 

\begin{enumerate}

\item To identify the phase space variables that allow writing the system of cosmological equations in the form of an autonomus system of ordinary differential equations (ODE), say: $$x_i=(x_1,x_2,...x_n)\;.$$

\item With the help of the chosen phase space variables, to build an autonomous system of ODE out of the original system of cosmological equations ($\tau$ is the time-ordering variable, not necessarily the cosmic time): $$\frac{dx_i}{d\tau}=f_i(x_1,x_2,...x_n)\;.$$ Notice that the RHS of these equations do not depend explicitly on $\tau$ (that is the reason why the system is called autonomous).

\item To identify the phase space spanned by the chosen variables $(x_1,x_2,...x_n)$, that is relevant to the cosmological model under study. This amounts, basically, to define the range of the phase space variables that is appropriate to the problem at hand: $$\Psi=\{(x_1,x_2,...x_n):\text{bounds on the}\;x_i\text{-s}\}\;.$$ 

\item Finding the equilibrium points of the autonomous system of ODE, amounts to solve the following system of algebraic equations on $(x_1,x_2,...x_n)$: $$f_i(x_1,x_2,...x_n)=0\;.$$

\item Next one linearly expands the equations of the autonomous system of ODE in the neighborhood of the equilibrium points $\bar p_k=p_k(\bar x_1,\bar x_2,...\bar x_n)$, $k=1,2,...m$: I. e., one replaces $x_i\rightarrow\bar x_i+e_i$, where $e_i$ are the small (linear) perturbations around the equilibrium points. Hence the system of ODE becomes a system of linear equations to determine the evolution of the $e_i$-s: $$\frac{de_i}{d\tau}=\bar f_i+\sum_{j=1}^n\left(\frac{\partial f_i}{\partial x_j}\right)_{\bar p}e_j+{\cal O}(e_i^2)\;,$$ otherwise, since $\bar f_i=f_i(\bar p)=0$, then $$\frac{de_i}{d\tau}=\sum_j^n[M(\bar p)_i^j]\;e_j+{\cal O}(e_i^2)\;,$$ where we have introduced the linearization or Jacobian matrix $[M^j_i]=\partial f_i/\partial x_j$.

\item The next step is to solve the secular equation to determine the eigenvalues $\lambda_i$ of the linearization matrix at the given equilibrium point $\bar p$: $$\det|M(\bar p)^j_i-\lambda\;U^j_i|=0\;,$$ where $[U^j_i]$ is the unit matrix. 

\item Once the eigenvalues of the linearization around a given equilibrium point $\bar p$ have been computed, the evolution of the perturbations is given by $$e_i(\tau)=\sum_j^n (e_0)_i^j\exp{(\lambda_j\tau)}\;,$$ where the amplitudes $(e_0)_i^j$ are constants of integration.

\end{enumerate} If all of the eigenvalues have negative real parts, the perturbations decay with $\tau$, i. e., the equilibrium point is stable against linear perturabtions. The corresponding equilibrium point is said to be a future attractor. If at least one of the eigenvalues has positive real part, the perturbations grow with $\tau$ so that these are not stable in the direction spanned by the given eigenvalue. Hence the point is said to be a saddle. The perturbations around a given equilibrium point are unstable, in other words the point is a past attractor (a source point in the phase space), if all of the eigenvalues have positive real parts. Points whose linearization is characterized by complex eigenvalues are said to be spiral equilibrium points, and are commonly associated with oscillatory behavior of the corresponding solution. If at least one of the eigenvalues has a vanishing real part, the equilibrium point is said to be non-hyperbolic. In the latter case, in general, and unless some of the non-vanishing real parts of the eigenvalues are of opposite sign, one can not give conclusive arguments on the stability of the equilibrium point. Other techniques have to be applied.

\end{document}